\documentclass[11pt]{article}
\usepackage[english]{babel}
\input{epsf}
\usepackage{mathtext}
\usepackage{graphics,dcolumn,bm,float}
\usepackage{amssymb,amsmath,rotate,color}
\usepackage{times}
\newcommand{\CA}{{\cal A}}
\newcommand{\CB}{{\cal B}}
\newcommand{\CC}{{\cal C}}
\newcommand{\CD}{{\cal D}}
\newcommand{\CH}{{\cal H}}

\newcommand{\be}{\begin{equation}}
\newcommand{\ene}{\end{equation}}
\newcommand{\ba}{\begin{array}}
\newcommand{\ea}{\end{array}}

\newcommand{\bdelta}{\mbox{\boldmath$\delta$}}

\begin{document}

\title{Coherent electron dynamics in monolayer $\text{MoS}_2$ under ultrashort optical pulse}
\author{Sina Pashalou, and Hadi Goudarzi\footnote{Corresponding author; e-mail address: h.goudarzi@urmia.ac.ir}\\
\footnotesize\textit{Department of Physics, Faculty of Science, Urmia University, P.O.Box: 165, Urmia, Iran}}
\date{}
\maketitle

\begin{abstract}
A theoretical investigation on the coherent Dirac-like quasiparticle dynamics in monolayer $\text{MoS}_2$ under an ultrashort optical pulse irradiation is presented. Particularly, we remain specific features of ML-MDS associated with the mass asymmetry $\alpha$ and topological aspect $\beta$ parameters resulting in Schr$\stackrel{..}{\text{o}}$dinger type wavevector of charge carriers. The direct band gap, spin-resolved valence band spilitting and valley degeneracy breaking due to strong spin-orbit coupling affect ultrafast dynamics of Dirac fermions. Because the duration of the optical pulse is less than the electron scattering time, which is $\sim 10-100\ fs$, the electron dynamics in electric field of the optical pulse is coherent, and consequently, we can describe the coupling of electron with strong electromagnetic field by the time-dependent Schr$\stackrel{..}{\text{o}}$dinger equation.
The conduction band and valence band coupling via the strong electric field of pulse ($0.2-2.5 \ V/\text{\AA}$) gives rise to appearance of a dipole moment during the pulse is applied. We find that the dipole is complex, originating from the existence of band gap. We show an asymmetric singularities in Dirac points for absolute of dipole moment. We solve the resulting coupled evolution equations for the expansion coefficient of valence and conduction bands to obtain probability of population transition between valence and conduction bands.
The irreversible electron dynamics as a key feature of two-dimensional Dirac matters in interacting with an ultrashort pulse strongly depends on the electronic structure of $\text{MoS}_2$.
Furthermore, we present conduction band population distribution with an asymmetric exhibition at the each pair of Dirac points ($K$ and $K'$), when the pulse ends. This leads to valley polarization effect.
The forced electric current is transferred on the surface of $\text{ML-MDS}$ in the pulse field direction, due to irreversible dynamics. These results can convince possibility to introduce device-friendly optoelectronic applications for $\text{MoS}_2$.
\end{abstract}
\textbf{keywords}: monolayer $\text{MoS}_{2}$; femtosecond laser pulse; dipole matrix element; conduction-band population; optoelectronics

\section{INTRODUCTION}

The coherent electron dynamics in semiconductors under optical pulse irradiation is of theoretical and experimental great interest \cite{1,2,3,4,5,6,7,8}, considering the fundamental effects such as transport properties of charge carries, conduction band (CB) population transition and optical absorption properties \cite{9,10}. The strong and ultrashort optical pulse with just a few oscillations effectively affects the electron dynamics in a condensed matter by its perpendicular or in-plane component of electric field \cite{6,11,12}. Such phenomena strongly depend on the dynamical properties of charge carriers in the structure. The coherent condition implies that the duration of pulse is less than the electron-electron scattering time. Then, necessarily, the pulse duration should be on femtosecond scale. Among the solids, two-dimensional Dirac-like materials, with pioneering atomic layer graphene \cite{13,14}, present distinct response to such an optical pulse induction, comparing with the ordinary semiconductors. In insulators, the reversibility of electron dynamics in the presence of an optical pulse has been investigated \cite{15,16}. Interaction of electrons in metals with an electromagnetic field gives rise to high frequency Bloch oscillation \cite{17}. Particularly in graphene, the CB population caused by strong electric field of laser pulse is irreversible after the pulse ends \cite{18,19}. In silicene and germanene \cite{20}, as graphene-like materials, subjected to a femtosecond strong optical pulse, it is observed that new exhibitions such as optical rectification and generation of electric current in both parallel and normal to the in-plane field direction are achieved.
 
In this paper, we theoretically study the coherent electron dynamics in another close to same structure families of 2D Dirac-like materials, which is transition metal dichalcogenide ($\text{TMD}$), specifically monolayer molybdenum disulfide ($\text{ML-MDS}$) under irradiation of ultrafast optical pulse. The main difference between monolayer $\text{MoS}_{2}$ in comparing to graphene is that $\text{ML-MDS}$ exhibits a stable charge exciton state even at room temperature, a property that is desirable for various optoelectronic and photonic applications \cite{21,22,23,24,25,26,27}, its band structure demonstrates strong spin-orbit coupling ($\text{SOC}$) of order $\sim 80\ meV$. Optical band gap between conduction and valence bands and, importantly, valley-degeneracy breaking. The hexagonal lattice of $\text{ML-MDS}$ consists of two-type heavy atom $\text{Mo}$ and $\text{S}$ with breaking inversion symmetry leading to hexagonal band structure in Brillouin zone, as shown in Figs. 1 and 3. The valence band $\text{SOC}$-resolved sizable splitting is another significant feature of $\text{ML-MDS}$, which can play a role in population transition to the conduction band forced by electric field of optical pulse. In particular, the topological corrections for $\text{ML-MDS}$, associated to the mass-difference between electron and hole \cite{28}, which are along with the Schr$\stackrel{..}{\text{o}}$dinger-type wave vector, can affect symmetry of charge transport.

Recently, an experimentally probe and also a simulation method to investigate the effect of intense femtosecond laser excitation on the bulk and monolayer $\text{MoS}_{2}$ have been presented \cite{29}. Another very recently experimental investigation on the optical susceptibility of $\text{ML-MDS}$ reveals particular feature of $\text{MoS}_{2}$, such that electron-electron coulomb interaction belonging to different valleys $\text{K}$ and $\text{K}^{\prime}$ is comparable to the two-body coulomb interaction between two electrons close together in phase space \cite{30}. In \cite{30}, authors showed that full spin polarization of electrons is achieved in the presence of a magnetic field, despite the Mermin-Wagner theorem . Therefore, it seems that further theoretical studies on the evaluating electron dynamics in 2D Dirac-like materials interacting with optical pulse is worthy. In Ref. \cite{31}, authors showed a fundamental possibility to occur a significant CB population and valley polarization in $\text{TMD}$ during just one period of a chiral laser field.

We focus on the interband transition in monolayer $\text{MoS}_{2}$ interacting with an ultrashort laser pulse. The outline of our paper is the following. In Sec. $2$, we present proposed structure and related formalism to obtain the explicit form of dipole matrix elements and exact relation for time-dependent conduction band population (CBP). We also represents the electric current, which is resulted from interband transition of electrons in time-dependent electric field and charge transfer through the system. Also Sec. $2$ is devoted to solve the set of coupled differential equations, simultaneously to obtain time-dependent expansion coefficients and sum over the all momentum in the first Brillouin zone. The numerical results and corresponding discussion are presented in Sec. $3$. Finally, a brief conclusion is given. 

\section{THEORETICAL MODEL}
\subsection{Time-dependent Hamiltonian of ML-MDS}

We consider an optical pulse, that is incident normally on a monolayer molybdenum disulfide with linear polarization in $\text{ML-MDS}$ plane, and parameterize it by the following single-oscillation form with time-dependent electric field:
\begin{equation}
F(t)={F}_{0}\ e^{-u^{2}}\left(1-2u^{2}\right).
\end{equation}
This optical field is an idealization of the actual $1.5$-oscillation pulses, which is used during last few years in experiments \cite{9,32}, and ${F}_{0}$ is the amplitude of pulse field, that is related to the pulse power $P=cF_{0}^{2}/4\pi$, $c$ is the speed of light, $u=t/\tau$ and $\tau$ denotes the pulse length, which is set $\tau=1\ fs$, corresponding to carrier frequency $\omega \sim 1.5\ eV/\hbar$. Note that, due to given pulse shape, the pulse has always zero area, $\int^{+\infty}_{-\infty}{F}(t)dt=0$, satisfying requirement the pulse to propagate in far-field zone. The plane of polarization is characterized by angle $\phi$ measured relative to axis $x$. Here, $x$ and $y$ coordinates system introduced in the plane of $\text{ML-MDS}$, and is determined by the crystallographic structure of $\text{ML-MDS}$, see Fig. 2. Similar to monoatomic layer graphene, the Brillouin zone of monolayer $\text{MoS}_{2}$ is hexagonal and around the edges of this zone the low energy fermionic excitations behave as massive Dirac particles. The two sublattices $\text{A}$ and $\text{B}$ include two different atoms $\text{Mo}$ and $\text{S}$. The presence of heavy transition metal atom $\text{Mo}$ gives rise to appearance of a sizable spin-orbit coupling $\sim 80\ meV$. Consequently, unlike graphene, we meet with two distinct dynamical features; breaking the valley degeneracy and valence-band spin-resolved splitting. Therefore, the electronic structure of $\text{ML-MDS}$ exhibits a valley degree of freedom indicating that the valence and conduction bands consist of two non-degenerate valleys ($\text{K}$ and $\text{K}^{\prime}$) located at the corners of the hexagonal Brillouin zone as shown in Figs. 3(b) and (c).

These characteristics can actually cause a distinct coherent electron dynamics interacting with electric field of a femtosecond laser pulse. The coherency may be presented by the fact that we take the characteristic electron-electron scattering time in $\text{ML-MDS}$ to be similar to the corresponding time in graphene, $10-100\ fs$, \cite{33,34}. Therefore, by tuning the duration of the optical pulse in our proposed system, the coherent dynamics can lead to electron transition from valence band to conduction band between different $\text{K}$ and $\text{K}^{\prime}$ valleys.

The Hamiltonian of electron in $\text{ML-MDS}$ interacting with time-dependent electric field has the form:
\begin{equation}
\CH(t)=H_{0}+e\textbf{F}(t)\cdot\textbf{r},
\end{equation}
where $H_{0}$ is the field-free electron Hamiltonian, $\textbf{r}=(x,y)$ is a two dimensional vector and the electric field components are $\textbf{F}(t)=(\text{F}(t)\cos\phi,\text{F}(t)\sin\phi)$. Below, we consider the case of $\phi=0$ i.e, pulse is polarized along the $x$ axis. We consider a nearest-neighbor tight-binding model, which describes coupling between the two atoms $\text{S}$ and $\text{Mo}$ of $\text{ML-MDS}$ with coupling constant $\gamma\cong 1.68\ eV$.
The $\text{ML-MDS}$ Hamiltonian $H_{0}$ is introduced within the nearest-neighbor tight-binding model by a $2\times 2$ matrix:
\begin{equation}
H_{0}=
\begin{pmatrix}
\Delta+\frac{\hbar^{2}|k|^{2}}{4m_{0}}(\alpha+\beta)-E_{f}&\gamma f(\textbf{k})\\ \gamma f^{\ast}(\textbf{k})&\lambda s\tau-\Delta+\frac{\hbar^{2}|k|^{2}}{4m_{0}}(\alpha-\beta)-E_{f}
\end{pmatrix},
\end{equation}
where $E_{f}$ is Fermi energy. $\Delta/2\cong 0.95\ eV$ is the direct band gap between conduction and valence bands and $\lambda\cong 80\  meV$ and $s=\pm$ denote the spin-orbit coupling and spin index (spin up and down), respectively. The valley index $\tau=\pm$ indicates the $\text{K}$ and $\text{K}^{\prime}$ valleys. The bare electron mass is $m_{0}$, and two numeric topological parameters are evaluated by $\alpha=\frac{m_{0}}{m_{+}}$ and $\beta=\frac{m_{0}}{m_{-}}-\frac{8m_{0}v^{2}_{F}}{\Delta-\lambda}$, where $m_{\pm}=\frac{m_{e}m_{h}}{m_{h}\pm m_{e}}$. $m_{e}$ and $m_{h}$ denote the mass of electron and hole \cite{28}. The difference between $m_{e}$ and $m_{h}$ leads to appearance mass difference and topological parameters, $\alpha=0.43$ and $\beta=2.21$, respectively, associated with the particular characteristic features of $\text{ML-MDS}$. These parameters are accompanied by the Schr$\stackrel{..}{\text{o}}$dinger type of quasi-particle wave vector, and to precisely investigate, we proceed to take into account the $\alpha$ and $\beta$ parameters contribution in the present system. $v_{f}\cong 0.53\times 10^{6}\ ms^{-1}$ is Fermi velocity of charge carriers. The off-diagonal term of Hamiltonian (3)
\begin{equation}	f(\textbf{k})=\exp{\left[i\left(\textbf{k}\cdot\bdelta_{\mathbf{1}}\right)\right]}+\exp{\left[i\left(\textbf{k}\cdot\bdelta_{\mathbf{2}}+\omega\right)\right]}+\exp{\left[i\left(\textbf{k}\cdot\bdelta_{\mathbf{3}}-\omega\right)\right]}=|f(\textbf{k})|e^{i\phi_{\textbf{\textbf{k}}}}.
\end{equation}
is the structure factor with $\omega=\frac{2\pi}{3}$. In-plane momentum is ${\textbf{k}}=({k}_{x},{k}_{y})$ and $\bdelta (i=1,2,3)$ is the in-plane components of the lattice vectors $\delta_{i\pm}$. The subindices of the hopping matrices indicate the nearest-neighbor vectors (as shown in Fig. 2),
 $$\delta_{1\pm}=a\left(\frac{\sqrt{3}}{2}\cos\theta,-\frac{1}{2}\cos\theta,\pm \sin\theta\right),$$
 $$\delta_{2\pm}=a\left(0,\cos\theta,\pm \sin\theta\right),$$
 $$\delta_{3\pm}=a\left(\frac{-\sqrt{3}}{2}\cos\theta,-\frac{1}{2}\cos\theta,\pm \sin\theta\right).$$  
where the lattice constant $a=2.43 \text{\AA}$ and $\theta=40.7^{\circ}$ are the $\text{Mo-S}$ bond length and the angle between the bond length and $\text{Mo}$'s plane, respectively. We consider no longer the $z$-components of ($\delta_{\pm z}$), because it is  canceled.
  
The energy spectrum of Hamiltonian $H_{0}$ consists of conduction band ($\pi^{\ast}$) and valence band ($\pi$) with the dispersion energy:
\begin{equation}
\varepsilon_{c,v}(\textbf{k})=-E_{f}+\frac{\lambda s \tau}{2}+\frac{\hbar^{2}|k|^{2}}{4m_{0}}\alpha\pm \sqrt{\left(\Delta-\frac{\lambda s \tau}{2}+\frac{\hbar^{2}|k|^{2}}{4m_{0}}\beta\right)^{2}+\gamma^{2}|f(\textbf{k})|^2}.
\end{equation}
Using the eigenvalue equation along with normalization condition, we arrive at the corresponding wave functions as:
\begin{equation}
\psi^{(c)}_{\textbf{k}}=\frac{e^{i{\textbf{k}}\cdot{\textbf{r}}}}{\sqrt{2}}
\begin{pmatrix}
\sqrt{\frac{\CA}{\CC}}\ e^{i\phi_{\textbf{k}}} \\ \sqrt{\frac{\CB}{\CC}}
\end{pmatrix},
\end{equation}
 and
\begin{equation}
\psi^{(v)}_{\textbf{k}}=\frac{e^{i{\textbf{k}}\cdot{\textbf{r}}}}{\sqrt{2}}
\begin{pmatrix}
-\sqrt{\frac{\CB}{\CC}}\ e^{i\phi_{\textbf{k}}} \\ \sqrt{\frac{\CA}{\CC}}
\end{pmatrix},
\end{equation}
where we have defined:
$$\CA= \varepsilon_{c}-\lambda s \tau+\Delta-\frac{\hbar^{2}|k|^{2}}{4m_{0}}(\alpha-\beta)+E_{f},$$
$$\CB= \varepsilon_{c}-\Delta-\frac{\hbar^{2}|k|^{2}}{4m_{0}}(\alpha+\beta)+E_{f},$$
$$\CC= \varepsilon_{c}-\frac{\lambda s \tau}{2}-\frac{\hbar^{2}|k|^{2}}{4m_{0}}\alpha+E_{f}.$$

\subsection{Transition probability}

When the duration of the laser pulse is less than the characteristic electron scattering time, the coherent transition of electron in external electric field of the optical pulse can be described by the time-dependent Schr$\stackrel{..}{\text{o}}$dinger equation:
\begin{equation}
i\hbar\frac{\partial{\psi}}{\partial{t}}=\CH(t)\psi.
\end{equation}
The electric field of optical pulse accelerates the electrons of $\text{ML-MDS}$ in the direction of the field polarization through the $\text{ML-MDS}$ plane, and also change the wave vector $\textbf{k}$ of electrons. Moreover, the electric field can affect both intraband and interband electron dynamics, in which the interband electron dynamics gives rise to coupling the conduction band (CB) and valence band (VB) states, resulting in a redistribution of electrons between two bands.

In the reciprocal space, the electron dynamics is described by acceleration theorem:
\begin{equation}
\hbar\frac{d\textbf{k}}{dt}=e\textbf{F}(t).
\end{equation}
For an electron with initial momentum $\textbf{q}$, the electron dynamics is described by the time-dependent wave vector, which is given by the solution of Eq. ($9$) :
\begin{equation}
\textbf{k}_{T}(\textbf{q},t)=\textbf{q}+\frac{e}{\hbar}\int^{t}_{-\infty} \textbf{F}\left(t_{1}\right)dt_{1},
\end{equation}
and the corresponding wave functions are the Houston functions \cite{35}:
\begin{equation}
\phi^{H}_{\zeta \textbf{q}}(\textbf{r},t)=\psi^{\zeta}_{\textbf{k}_{T}(\textbf{q},t)}(\textbf{r})\exp\left[-i\int^{t}_{-\infty}dt_{1} \varepsilon_{\zeta}\left[{\textbf{k}_{T}(\textbf{q},t_{1})}\right]\right],
\end{equation}
where $\zeta=v$ for valence band and $\zeta=c$ for conduction band. We express the general solution of the time-dependent Schr$\stackrel{..}{\text{o}}$dinger equation ($8$) by the superposition of Houston function:
\begin{equation}
\psi_{\textbf{q}}(\textbf{r},t)=\sum_{\zeta=c,v}\beta_{\zeta \textbf{q}}(t)\phi^{H}_{\zeta \textbf{q}}(\textbf{r},t),
\end{equation}
where $\beta_{\zeta \textbf{q}}(t)$ is corresponding time-dependent expansion coefficients. Coupling of CB and VB states via the external electric field is determined by an interband dipole matrix element. The interband dipole matrix elements is diagonal in reciprocal space, so that the states not coupled by the pulse field. Such coupling of states with the same value of $\textbf{q}$ is the property of the coherent dynamics. For incoherent dynamics , the electron scattering couples the states with different wave vectors $\textbf{q}$. In this case the dynamics can be  described by the density matrix. Substituting wave functions expressed in Eq. ($12$) into the time-dependent Schr$\stackrel{..}{\text{o}}$dinger equation results in two decoupled equations.

Finally, one can find, the expansion coefficients satisfy the following system of differential equations:
\begin{equation*}
\frac{d\beta_{c\textbf{q}}(t)}{dt}=-i\frac{\textbf{F}(t)\textbf{Q}_{\textbf{q}}(t)}{\hbar}\beta_{v\textbf{q}}(t),
\end{equation*}
\begin{equation}
\frac{d\beta_{v\textbf{q}}(t)}{dt}=-i\frac{\textbf{F}(t)\textbf{Q}^{\ast}_{\textbf{q}}(t)}{\hbar}\beta_{c\textbf{q}}(t),
\end{equation}
where the vector function $\textbf{Q}_{\textbf{q}}(t)$ is proportional to the interband dipole matrix element:
\begin{equation}	\textbf{Q}_{\textbf{q}}(t)=\textbf{D}\left[\textbf{k}_{T}(\textbf{q},t)\right]\exp\left[-\frac{i}{\hbar}\int^{t}_{-\infty}dt_{1}({\varepsilon_{c}\left[\textbf{k}_{T}(\textbf{q},t_{1})\right]-\varepsilon_{v}\left[\textbf{k}_{T}(\textbf{q},t_{1})\right]})\right],
\end{equation}
where $\textbf{D}(\textbf{k})\equiv\left[D_{x}(\textbf{k}),D_{y}(\textbf{k})\right]$ is the dipole matrix element between the states of the conduction and valence bands with wave vector $\textbf{k}$,
\begin{equation}
\textbf{D}(\textbf{k})=\left\langle \psi^{c}_{\textbf{k}}|e\textbf{r}| \psi^{v}_{\textbf{k}}\right\rangle.
\end{equation}
Substituting the conduction and valence band wave-functions ($6$) and ($7$) into Eq. ($15$), we arrive at the following expressions for the interband dipole matrix elements :
\begin{equation}
D_{j}(\textbf{k})=\sqrt{1-\left({\frac{\CD}{\CC}}\right)^{2}}
\ Z_{j}-\frac{ie}{2}\frac{\Lambda}{\CC\sqrt{\CC^{2}-\CD^{2}}} \ ;\ \ \ \ (j\equiv x,y),
\end{equation}
where we have :
$$\CD=\Delta-\frac{\lambda s \tau}{2}+\frac{\hbar^{2}|k|^{2}}{4m_{0}}\beta,$$
$$\Lambda=\frac{\partial{\varepsilon_{c}}}{\partial{k_{j}}}\CD+\frac{\hbar^{2}k_{j}}{2m_{0}}\left[\left(\frac{\lambda s \tau}{2}-\Delta\right)\alpha+\left(\frac{\lambda s \tau}{2}-\varepsilon_{c}-E_{f}\right)\beta\right],$$
\begin{equation*}	\text{Z}_{x}=\frac{ae\sqrt{3}\cos\theta}{4}\frac{\cos(\text{X}-\text{Y})-\cos(\text{Y}-\text{Z})}{3+2[\cos(\text{X}-\text{Y})+\cos(\text{Y}-\text{Z})+\cos(\text{X}-\text{Z})]},
\end{equation*}
\begin{equation*}	\text{Z}_{y}=\frac{ae\cos\theta}{4}\frac{\cos(\text{X}-\text{Y})+\cos(\text{Y}-\text{Z})-2\cos(\text{X}-\text{Z})}{3+2[\cos(\text{X}-\text{Y})+\cos(\text{Y}-\text{Z})+\cos(\text{X}-\text{Z})]}.
\end{equation*}
and
$$\text{X}-\text{Y}=\frac{a\sqrt{3}}{2}\cos\theta \ k_{x}-\frac{3a}{2}\cos\theta \ k_{y}-\omega,$$ 
$$\text{Y}-\text{Z}=\frac{a\sqrt{3}}{2}\cos\theta \ k_{x}+\frac{3a}{2}\cos\theta \ k_{y}+2\omega,$$
$$\text{X}-\text{Z}=a\sqrt{3}\cos\theta \ k_{x}+\omega.$$
The system of equations ($13$) describes the interband electron dynamics and determines the mixing of the CB and the VB states in the electric field of the pulse. There are two solutions of the system ($13$), which correspond to two initial conditions: $(\beta_{v\textbf{q}},\beta_{c\textbf{q}})=(1,0)$ and $(\beta_{v\textbf{q}},\beta_{c\textbf{q}})=(0,1)$. These solutions determine the evolution of the states, which are initially in the VB or CB, respectively. The mixing of the states from different bands and the dynamics of an electron which initially lies in the VB is characterized by the time-dependent expansion coefficient $|\beta_{c\textbf{q}}(t)|^{2}$. The time-dependent total transition of CB is expressed by the following expression :
\begin{equation}
N_{CB}(t)=\sum_{\textbf{q}}|\beta_{c\textbf{q}}(t)|^{2},
\end{equation}
where the sum is over all momentum in the first Brillouin zone.

Furthermore, we proceed to evaluate the electric current, which is resulted from interband redistribution of electrons in time-dependent electric field. Electric current can be calculated in terms of the operator of velocity :
\begin{equation}
	J_{i}(t)=\frac{e}{a^2}\sum_{\textbf{q}}\sum_{\zeta_{1}=v,c}\sum_{\zeta_{2}=v,c}\beta_{\zeta_{1}\textbf{q}}^{\ast}(t)\emph{V}_{j}^{\ \zeta_{1}\zeta_{2}}(t) \beta_{\zeta_{2}\textbf{q}}(t).
\end{equation}
$\emph{V}_{j}^{\ \zeta_{1}\zeta_{2}}$ is the matrix elements of the velocity operator $\hat{\emph{V}}_{j}=\frac{1}{\hbar}\frac{\partial{H_{0}}}{\partial{k_{j}}}$. Using wave functions of the conduction and valence bands, the matrix elements of the velocity operator is given by :
\begin{equation}
\emph{V}_{j}^{\ \zeta_{1}\zeta_{2}}=\left\langle \psi_{\textbf{k}}^{\zeta_{1}}|\hat{\emph{V}}_{j}|\psi_{\textbf{k}}^{\zeta_{2}}\right\rangle.
\end{equation}
Substituting the conduction and valence bands wave functions (6) and (7) into Eq. (19), we obtain the following expressions for the matrix elements of the velocity operator :
\begin{equation*}
\emph{V}_{x}^{cc}=\Gamma^{+}_{x}-\sqrt{3}\rho\Sigma\Omega \ , \ \ \emph{V}_{x}^{vv}=\Gamma^{-}_{x}+\sqrt{3}\rho\Sigma\Omega,
\end{equation*}
\begin{equation*}
\emph{V}_{y}^{cc}=\Gamma^{+}_{y}+\rho\Sigma\Xi \ , \ \ \emph{V}_{y}^{vv}=\Gamma^{-}_{y}-\rho\Sigma\Xi,
\end{equation*}
and
\begin{equation*}
\emph{V}_{x}^{cv}=\frac{-\hbar k_{x}}{2 m_{0}}\beta\Sigma+\sqrt{3}\rho\left[i\Theta-\left(\frac{\CD}{\CC}\right)\Omega\right],
\end{equation*}
\begin{equation*}
\emph{V}_{y}^{cv}=\frac{-\hbar k_{y}}{2 m_{0}}\beta\Sigma+\rho\left[i\Phi+\left(\frac{\CD}{\CC}\right)\ \Xi\right],
\end{equation*}
where we have defined:
$$
\Gamma^{\pm}_{x(y)}=\frac{\hbar k_{x(y)}}{2 m_{0}}\alpha\pm\frac{\hbar k_{x(y)}}{2 m_{0}}\beta\left(\frac{\CD}{\CC}\right),
$$
$$
\rho=\frac{a\gamma\cos\theta}{2\hbar} , \ \ \Sigma=\sqrt{1-\left(\frac{\CD}{\CC}\right)^{2}} \ ,
$$
$$
\Omega=\sin(\textbf{k}\cdot\bdelta_{\mathbf{1}}-\phi_{\textbf{k}})-\sin(\textbf{k}\cdot\bdelta_{\mathbf{3}}-\omega-\phi_{\textbf{k}})\ ,$$
$$\Xi=\sin(\textbf{k}\cdot\bdelta_{\mathbf{1}}-\phi_{\textbf{k}})-2 \sin(\textbf{k}\cdot\bdelta_{\mathbf{2}}+\omega-\phi_{\textbf{k}})+\sin(\textbf{k}\cdot\bdelta_{\mathbf{3}}-\omega-\phi_{\textbf{k}}),$$
$$\Theta=\cos(\textbf{k}\cdot\bdelta_{\mathbf{1}}-\phi_{\textbf{k}})-\cos(\textbf{k}\cdot\bdelta_{\mathbf{3}}-\omega-\phi_{\textbf{k}})\ ,$$
$$\Phi=2\cos(\textbf{k}\cdot\bdelta_{\mathbf{2}}+\omega-\phi_{\textbf{k}})-\cos(\textbf{k}\cdot\bdelta_{\mathbf{1}}-\phi_{\textbf{k}})-\cos(\textbf{k}\cdot\bdelta_{\mathbf{3}}-\omega-\phi_{\textbf{k}}).$$
The interband matrix elements of the velocity operator, $\emph{V}_{x}^{cv}$ and $\emph{V}_{y}^{cv}$, are related to the interband dipole matrix elements \cite{36},
$$\emph{V}_{x}^{cv}=iD_{x}(\textbf{k})\frac{\left[\varepsilon_{c}(\textbf{k})-\varepsilon_{v}(\textbf{k})\right]}{\hbar},$$ and $$\emph{V}_{y}^{cv}=iD_{y}(\textbf{k})\frac{\left[\varepsilon_{c}(\textbf{k})-\varepsilon_{v}(\textbf{k})\right]}{\hbar}.$$

Note that within the nearest-neighbor tight-binding model, for the graphene we have electron-hole symmetry, which results in $\emph{V}_{y}^{cc}=-\emph{V}_{y}^{vv}$ \cite{18}. Whereas, in the case of $\text{ML-MDS}$, we find that  $\emph{V}_{y}^{cc}\neq-\emph{V}_{y}^{vv}$, which is in contrast with the graphene case. Further, we can calculate the charge transfer implemented by the in plane electric current. To this end, we consider that the generated current results in charge transfer through the system, which is obtained by :
$$
\emph{Q}_{tr,\mu}=\int_{-\infty}^{+\infty}dt\ J_{\mu}(t),
$$
where $\mu=||$ or $\bot$, which corresponds to the charge transfer along the direction of polarization of the laser pulse and in the direction perpendicular to polarization of the pulse, respectively. The transformed charge can be nonzero, whether the dynamics for the entire system (when pulse ends) is irreversible. For completely reversible dynamics, when the system returns to its initial state, the transformed charge may be exactly zero. Actually the residual population of $|\beta_{c\textbf{q}}(t)|^{2}$ determines reversibility of dynamics.
 
\section{NUMERICAL RESULTS AND DISCUSSION}

In this section, we proceed to investigate in detail the Dirac fermions dynamics of $\text{ML-MDS}$ interacting with a strong femtosecond laser pulse. Our first investigation is focused on analyzing the effect of intrinsic massive Dirac gap ($\Delta$), strong spin-orbit coupling ($\lambda_{SO}$) and ,in particular, asymmetry mass parameter $\alpha$ and topological related parameter $\beta$ on the dipole matrix element, which can connect two charges $e$ from VB and CB. According Eq. (16), the interband dipole matrix elements $D_{x}(\textbf{k})$ and $D_{y}(\textbf{k})$ are found to be complex functions, and strongly depend on the electron wave vector $\textbf{k}$. The absolute of dipole matrix elements are singular at the Dirac points in the corners of Brillouin zone, as shown in Figs. 4(a-c). Obviously, there is no coupling at the center of Brillouin zone. Fig. 4(a) shows that an asymmetric behavior is found for absolute dipole moment for any pair of $\text{K}$-$\text{K}^{\prime}$ Dirac points. This can be originated from the fact that valley degeneracy breaks in $\text{ML-MDS}$. The effect of $\text{SOC}$ on the dipole matrix element is very mild, near the critical point $\lambda_{\text{SO}}=0.08\ eV$, as shown in Fig. 4(c), since $\text{SOC}$ actually causes tuning the amount of splitting between VB and CB to be closer or farther with spin up or spin down polarization. Therefore, the height of singular peaks of dipole moment are slightly shortened. Whereas, as seen from Fig. 4(b) vanishing particularly $\beta$ parameter has a remarkable effect on dipole moment, that the inequality of moments singularity in Dirac points is intensified.

Next, we proceed to explore the key result of our proposed system, that is CBP. To study the evolution of the Dirac fermions dynamics in $\text{ML-MDS}$ under electric field of optical ultrafast pulse induction, we solve the set of two coupled differential equations (13), simultaneously. We set initial condition as $\left(\beta_{v\textbf{q}},\beta_{c\textbf{q}}\right)=(0,1)$, which denotes the condition that before interaction of electric field with quasiparticles, all states of the VB are occupied, and consequently, all states of CB are empty. First, we plot CBP as function of time for a pulse duration being $(1-2)$-oscillations, demonstrated in Fig. 5(a) for various electric field strengths. Remarkably, the irreversible electron dynamics is achieved for CBP, as a key feature of coherent dynamics of Dirac-like electrons in two dimensional materials. In contrast with previous works \cite{18,19}, the maximum of CBP occurs in its residue, when the pulse ends. It seems this result to be more acceptable. As expected, the residue of CBP increases with the increase of electric field strength of optical pulse. In Fig. 5(b), the effect of $\text{ML-MDS}$ characteristic particular dynamical parameters on CBP is presented. We show that the $\text{SOC}$ and electron-hole mass asymmetry parameters have no impact in the CB transition, whereas topological related parameter $\beta$ considerably enhances CBP. Also, we find that, even in the presence of a high band gap, the irreversible CB transition is observed, see Fig. 5(b) for $\Delta=1.9 \ eV$ panel. In Fig. 5(c), we represent dependence of CBP probability on the pulse incidence angle, where CBP enhances with the increase of angle. Indeed, in this case, the acceleration of charge carriers is realized to vary in in parallel or perpendicular direction in $\text{ML-MDS}$ plan during the pulse elapsed time. This influences conduction-valence band coupling via the accelerated electrons in various directions.

We also investigate distribution of electrons around Dirac points ($\text{K}$ and $\text{K}^{\prime}$) when the pulse ends by utilizing the acceleration theorem. Residual CB population distributions $\text{N}_{CB}(t)$ of the electrons in the reciprocal space for various field amplitude $F_{0}$ and incidence angle $\phi=0$ are illustrated in Figs. 6(a-c). After the pulse ends, excited electrons to CB determine redistribution of electrons at six point of corners of first Brillouin zone. We see that this CB distribution exists and, consequently, there is a strong interband coupling at Dirac points even in the presence of a sizable energy gap in \text{ML-MDS}. In Fig. (6) panels (a),(b) and (c), the distribution of the residual CBP for three different values of $F_{0}=0.2, 0.5, 1.0\ V/\text{\AA}$ are demonstrated. The electron distribution in the corners of Brillouin zone increases, and also becomes asymmetric with respect to specially $\textbf{k}_{x}$ axis with the increase of strength of pulse. This is in accordance to the corresponding asymmetry in absolute of dipole moment of structure. However, the asymmetric distribution is observed in a pair of $\text{K}$-$\text{K}^{\prime}$ valley, as shown in Fig. 6(c), that right side of the Dirac points are more populated than the left sides. This can be realized by the appearance of nonequivalent valleys in $\text{ML-MDS}$.

We proceed to reveal the signature of SOC and the size energy gap in the electron distribution. From Fig. 6(d), it is deduced that spin-orbit coupling as an important feature of $\text{ML-MDS}$ has no considerable affect on CB transition in Dirac points, since from electric field interaction view point, the presence of SOC in monolayer $\text{MoS}_{2}$ causes valence band displacement toward conduction band, and its values is negligible comparing to the size of band gap. Moreover, the size of band gap strongly influences population transition, where the intense of hot points is gained and new points of electron distribution emerge along the $\textbf{k}_{x}$ axis for constant $\textbf{k}_{y}$, when we set a smaller band gap, $\Delta=0.2\ eV$, as shown in Fig. 6(e). Finally, we plot the CB electric current in $x$-direction $\emph{V}_{x}^{cc}$ resulting from charge acceleration as a function of time in  this system, which is illustrated in Fig. 7. In the first half of the pulse, the forced current is low and negative, while in the second half denotes highly positive. Fig. 7 displays dynamics of the current, which is non-zero after the pulse ends, because of the asymmetric dipole moment and electron distribution in Dirac points. The current strongly depends on the strength of pulse electric field. 
  
\section{CONCLUSION}

In summary, we have demonstrated interband transition probability of Dirac fermions in \text{ML-MDS} in the presence of applying  an femtosecond laser pulse. In this case, because the duration of the optical pulse is less than the electron scattering time, the electron dynamics of system was considered to be coherent. The strong time-dependent electric field of optical pulse couples conduction and valence bands via the electric dipole moment implemented in Dirac points. Specifically, we have considered effect of direct band gap, spin-orbit coupling and topological characteristic $\alpha$ and $\beta$ parameters of \text{ML-MDS} on the dipole matrix elements singularity in the Dirac points and also on the conduction band population resulted from transition probability. Dipole matrix element of system has been found to be complex function, which is resulted from the existence of band gap. The numerical calculation of total electron transition rate in reciprocal space for all points of first Brillouin zone have shown weak dependence of CBP on the SOC for $0\leq\lambda_{SO}\leq 0.08\ eV$, whereas strong dependence on the direct band gap for $0\leq\Delta/2\leq 0.95\ eV$ and topology parameter $\beta=2.21$. The irreversible electron dynamics is obtained when the optical pulse is over, and the maximum CBP occurs in its residue. Redistribution of transferred electrons in conduction band has been presented for various strength of electric field of pulse. The effect of SOC and band gap of \text{ML-MDS} on the CB distribution in Dirac points has been illustrated. Importantly, we have found asymmetric distribution in two different Dirac points $K$ and $K'$. Therefore, our findings clearly reveals valley polarization with respect to the CB electron distribution, when a linear polarized pulse is used, which has been reported in \cite{31} for circular polarized pulse. However, the results presented can be experimentally verified by time-resolved angle-resolved photoemission spectroscopy (tr-ARPES) \cite{37,38}. The irreversible CB transition probability generates an electric current by the pulse force, which results in non-zero charge transfer through the system after the pulse ends. The asymmetric CB transition dynamics in Dirac points can be responsible to this behavior.

\flushleft{\textbf{Acknowledgment}}

The authors are grateful to M. Khezerlou for discussions in numerical results.

\newpage

\textbf{Figure captions}\\
\textbf{Figure 1} (Color online) Sketch of hexagonal lattice structure of monolayer $\text{MoS}_{2}$. The $\text{MoS}_{2}$ lattice consists of two sublattices $\text{A}$ and $\text{B}$ include two different atoms $\text{Mo}$ and $\text{S}$. The nearest-neighbor coupling, which is characterized by the hopping integral $\gamma$ is shown. Points $\text{K}$ and $\text{K}^{\prime}$ are two non-degenerate Dirac points, corresponding to two valleys. Blue line with arrows shows the time-dependent irradiated ultrashort pulse. The electric field of the optical pulse makes angle $\phi$ with normal direction on $\text{MoS}_{2}$ plane.\\
\textbf{Figure 2} Side and top views of the lattice structure of $\text{MoS}_{2}$ are seen, where \text{Mo} atom is surrounded by six \text{S} atoms.\\
\textbf{Figure 3(a), (b), (c)} (Color online) (a) The energy spectrum of \text{ML-MDS} as a function of wave vectors $k_x$ and $k_y$ with six Dirac points in reciprocal space. Red solid lines with arrow indicate (b) spin up and (c) spin down, which leads to valence band spin-splitting effect.\\
\textbf{Figure 4(a), (b), (c)} (Color online) (a) Plot of absolute of interband electric dipole moment $|\text{D}_{x}|$ as a function of wave vectors $k_x$ and $k_y$. The dipole matrix element is singular near the Dirac points $\text{K}$ and $\text{K}'$. The intensity of dipole moment decreases (b) strongly for $\alpha=\beta=0$, and (c) very low for $\lambda_{\text{SOC}}=0$.\\
\textbf{Figure 5(a), (b), (c)} (Color online) Conduction band population, $\text{N}_{\text{CB}}(t)$, as a function of time, calculated from Eq. (17), (a) for different amplitudes of electric field of optical pulse, which for all states we take pulse incidence angle $\phi=0$. The corresponding electric field, $F(t)$, of the laser pulse is also shown as a function of time (brown circe line). (b) The plots show the results of CBP for characteristic parameters of \text{ML-MDS}; $\alpha$, $\beta$, $\lambda$ and $\Delta$. (c) For $F_{0}=2\ V/\text{\AA}$, the curves show the results for various angles of incidence of pulse.\\
\textbf{Figure 6(a), (b), (c), (d), (e)} (Color online) Conduction band electron distribution in Dirac points after the pulse ends as a function of wave vector $k_x$ and $k_y$ for different amplitudes $F_0=0.2, 0.5, 1.0 \ V/\text{\AA}$ of optical pulse is presented in (a),(b) and (c) panels. Only the first Brillouin zone is shown. Effect of SOC $\lambda$ and band gap $\Delta$ on the electron distribution is shown in (d) for $\lambda_{SO}=0$ and (e) for $\Delta=0.2\ eV$.\\
\textbf{Figure 7} (Color online) CB electric current density forced by electric field of pulse in the $x$-direction as a function of time for three amount of amplitudes, $F_{0}=1, 2, 3\ V/\text{\AA}$.

\newpage

\end{document}